\documentclass[a4paper,12pt,english]{article}
\usepackage{amsfonts}
\usepackage{graphicx}
\usepackage{amsmath}
\usepackage{color}

\newcommand{\be}{\begin{equation}}
\newcommand{\ee}{\end{equation}}
\input{tcilatex}

\begin{document}


\begin{titlepage}
\begin{center}

\noindent{{\LARGE{On the timelike Liouville three-point function}}}

\smallskip
\smallskip

\smallskip
\smallskip
\smallskip
\smallskip
\noindent{\large{Gaston Giribet}}

\smallskip
\smallskip

\end{center}
\smallskip
\smallskip
\centerline{Physics Department, University of Buenos Aires FCEN-UBA and CONICET}
\centerline{{\it Ciudad Universitaria, Pabell\'on 1, 1428, Buenos Aires, Argentina.}}
\smallskip
\smallskip
\smallskip

\bigskip

\bigskip

\bigskip

\bigskip

\begin{abstract}

In a recent paper, D. Harlow, J. Maltz, and E. Witten showed that a particular proposal for the timelike
Liouville three-point function, originally due to Al. Zamolodchikov and to I. Kostov and V. Petkova, can actually be computed
by the original Liouville path integral evaluated on a new integration cycle. Here, we discuss a
Coulomb gas computation of the timelike three-point function and show that
an analytic extension of the Selberg type integral formulas involved reproduces the same expression,
including the adequate normalization. A notable difference with the spacelike calculation is pointed out.


\end{abstract}

\end{titlepage}


Liouville field theory finds its application in various areas of theoretical
physics, including string theory \cite{P}, three-dimensional general
relativity \cite{CvDH}, string theory in Anti-de Sitter space \cite{AdS3},
and supersymmetric gauge theory \cite{AGT}. The timelike version of
Liouville theory, on the other hand, has interesting applications as well.
For instance, timelike Liouville was considered in holographic quantum
cosmology \cite{FSSY}, in the study of tachyon condensation \cite{ST} and in
other time-dependent scenarios of string theory. In a recent paper \cite{HMW}
timelike Liouville theory was reexamined from a more general point of view.
After a thoughtful analysis of the analytic continuation of Liouville field
theory, the question was raised as to what extent the peculiar case of
timelike Liouville theory can actually be regarded as a conformal field
theory. A crucial ingredient in the discussion is the three-point
correlation function, which is computed in \cite{HMW} within the path
integral approach. The computation of this observable is a non-trivial
problem since, contrary to the naive expectation, the structure constants of
the timelike Liouville theory are not the analytic extension of the
structure constants of spacelike Liouville theory. In this note, with the
aim of contributing to the discussion, we perform a computation of the
timelike three-point function, alternative to that in \cite{HMW}, and show
that it reproduces the same result, including the adequate normalization. We
discuss the difference with respect to the spacelike three-point function
computation.

Let us begin by briefly reviewing Liouville theory. The action of Liouville
theory formulated on a closed manifold $\mathcal{C}$ is%
\begin{equation}
S_{\text{L}}[\varphi ,\mu ]=\frac{1}{4\pi }\int_{\mathcal{C}}d^{2}x\sqrt{g}%
\left( \sigma g^{ab}\partial _{a}\varphi \partial _{b}\varphi +Q_{\sigma
}R\varphi +4\pi \mu e^{2b\varphi }\right)  \label{SL}
\end{equation}%
where $b$ and $\mu $ are two real parameters, and $Q_{\sigma }=b+\sigma
b^{-1}$ with $\sigma =\pm 1$. The value $\sigma =+1$ corresponds to the
standard Liouville theory while $\sigma =-1$ corresponds to the theory with
the \textit{wrong sign} kinetic term. Using the string theory terminology,
we refer to these theories as the spacelike and the timelike models,
respectively. Action (\ref{SL}), at least in its spacelike version $\sigma
=+1$ that we understand better, defines a non-compact conformal field
theory. Primary operators of the theory are given by the exponential fields $%
V_{\alpha }(z)=e^{2\alpha \varphi (z)}$, which create states with conformal
dimension $\Delta _{\alpha }=\sigma $ $\alpha (Q_{\sigma }-\alpha )$. From
this, it is possible to check that the self-interaction term in (\ref{SL})
represents a marginal deformation. The central charge of the theory is given
by $c=1+6\sigma Q_{\sigma }^{2}$. Notice also that the timelike model $%
\sigma =-1$ can be alternatively obtained from the standard case $\sigma =+1$
by going to imaginary values of the parameter $b\rightarrow ib$ and the
Liouville field $\varphi \rightarrow -i\varphi $. This is, indeed, a
convenient way of thinking about the timelike model; however, here we prefer
to keep $\sigma $ in the formulae below and consider $b\in \mathbb{R}$.

The $n$-point correlation functions of local operators on a curve $\mathcal{C%
}$ are defined by%
\begin{equation}
\left\langle \tprod\nolimits_{i=1}^{n}V_{\alpha _{i}}(z_{i})\right\rangle _{%
\mathcal{C}}=\int_{^{\varphi _{\text{(}\mathcal{C}\text{)}}}}\mathcal{D}%
\varphi \ e^{-S_{\text{L}}[\varphi ,\mu
]}\tprod\nolimits_{i=1}^{n}e^{2\alpha _{i}\varphi (z_{i})}.
\end{equation}

Here we are interested in the three-point correlation functions on the
sphere. The three-point function defines the structure constants of the
theory, hereafter denoted as $C_{b}(\alpha _{1},\alpha _{2},\alpha _{3})$.
Namely%
\begin{equation}
\left\langle \tprod\nolimits_{i=1}^{3}V_{\alpha _{i}}(z_{i})\right\rangle _{%
\mathbb{S}^{2}}=\tprod\nolimits_{i<j}^{3}|z_{i}-z_{j}|^{\Delta _{ij}}\
C_{b}(\alpha _{1},\alpha _{2},\alpha _{3}),  \label{n}
\end{equation}%
with $\Delta _{ij}=\Delta _{\alpha _{1}}+\Delta _{\alpha _{2}}+\Delta
_{\alpha _{3}}-2\Delta _{\alpha _{i}}-2\Delta _{\alpha _{j}}$. It follows
that%
\begin{equation}
C_{b}(\alpha _{1},\alpha _{2},\alpha _{3})=\left\langle V_{\alpha
_{1}}(0)V_{\alpha _{2}}(1)V_{\alpha _{3}}(\infty )\right\rangle _{\mathbb{S}%
^{2}}=\int_{\varphi _{\text{(}\mathbb{CP}^{1}\text{)}}}\mathcal{D}\varphi \
e^{-S_{\text{L}}[\varphi ,\mu ]}\ e^{2\alpha _{1}\varphi (0)}e^{2\alpha
_{2}\varphi (1)}e^{2\alpha _{3}\varphi (\infty )}  \label{m}
\end{equation}%
where projective invariance is invoked and used to fix the three insertions
on the Riemann sphere, namely $z_{1}=0,$ $z_{2}=1,$ $z_{3}=\infty $. On the
Riemann sphere the field configurations are conditioned to obey the
asymptotic $\varphi (z)\sim -2Q_{\sigma }\log |z|$ for $|z|>>1$. In (\ref{m}%
), a\ factor $|z_{3}|^{-2\Delta _{3}}$ when taking the limit $%
z_{3}\rightarrow \infty $ was omitted for short.

The three-point function (\ref{n}) of spacelike Liouville theory has been
calculated by H. Dorn and H. Otto in Ref. \cite{DO} and independently by the
brothers Zamolodchikov in Ref. \cite{ZZ}; this is why the explicit formula
for $C_{b}(\alpha _{1},\alpha _{2},\alpha _{3})$ is usually referred to as
the DOZZ formula. Further details of the calculation of the structure
constants were given in Refs. \cite{T, Pakman}.

The solution of the timelike model has been early investigated by A.
Strominger and T. Takayanagi in \cite{ST} within the context of the closed
string theory tachyon dynamics. Afterwards, V. Schomerus in a beautiful
paper \cite{S} proposed the first satisfactory answer for the timelike
three-point function, namely for the structure constants of the theory with $%
c<1$ (which corresponds to non-real values of $b$ in the standard
Liouville.) The proposal of \cite{S} was obtained by taking the limit from
the expressions valid for $c\geq 25$ carefully. The limit is actually
delicate as the theory with $c\leq 1$ happens not to depend smoothly on $c$.
In fact, the timelike structure constants proposed in \cite{S} are not
analytic functions of the momenta $\alpha _{i}$, and the spacelike and
timelike theories cannot be related simply by Wick rotation analytically
continuing in the parameters $\alpha _{i}$ and $b$.

The analytic extension of Liouville three-point function to non-real values
of $b$ was also investigated in Refs. \cite{Z, 05, KP, KP2, McE}. In
particular, in \cite{Z} Al. Zamolodchikov proposed a timelike version of the
DOZZ formula which is not the naive analytic continuation of its spacelike
analogue. A similar proposal was independently given by I. Kostov and V.
Petkova \cite{05}. In a recent a paper by D. Harlow, J. Maltz, and E. Witten 
\cite{HMW}, it is showed that the timelike Liouville three-point function
proposed in \cite{Z} can be obtained by the original Liouville path integral
evaluated on a new integration cycle.

In this note, with the aim of contributing to the discussion of the timelike
model and in particular to the discussion of the timelike three-point
function, we will rederive this quantity using the Coulomb gas approach. The
Coulomb gas calculation of the $c<1$ three-point function was also discussed
in \cite{05} (see the discussion around Eq. (3.9) therein.) Here, we will
show that a natural analytic extension of the Selberg type integral formulas
involved reproduces the expression of \cite{Z} with the same normalization
as in \cite{HMW}. We will perform a detailed comparison between the
spacelike and timelike calculations and point out a notable difference. The
difference comes from a divergent factor that arises in the integration over
the zero-mode. While in the spacelike case the DOZZ formula is obtained by
considering the full correlation function, in the timelike case the
analogous formula proposed in \cite{Z} is reproduced by the residues
associate to resonant correlators, meaning that a simple pole has to be
extracted.

The Coulomb gas calculation of Liouville correlation functions was early
discussed in Ref. \cite{GL}. This consists in expanding the interaction term
of the Liouville action and performing the Wick contraction of the operators
using the free field theory. After integrating the zero mode $\varphi _{0}$
of the Liouville field, we can write (\ref{m}) as follows%
\begin{equation}
C_{b}^{(\sigma )}(\alpha _{1},\alpha _{2},\alpha _{3})=\Gamma (-s_{\sigma
})\mu ^{s_{\sigma }}b^{-1}\int_{\varphi _{\text{(}\mathbb{CP}^{1}\text{)}}}%
\mathcal{D}\widetilde{\varphi }\ e^{-S_{\text{L}}[\widetilde{\varphi },\mu
=0]}\ e^{2\alpha _{1}\widetilde{\varphi }(0)}e^{2\alpha _{2}\widetilde{%
\varphi }(1)}e^{2\alpha _{3}\widetilde{\varphi }(\infty
)}\tprod\nolimits_{r=1}^{s_{\sigma }}e^{2b\widetilde{\varphi }(w_{r})}
\label{TPF}
\end{equation}%
where $\widetilde{\varphi }=\varphi -\varphi _{0}$ are the fluctuation of
the field, and $s_{\pm }=b^{-1}(Q_{\pm }-\alpha _{1}-\alpha _{2}-\alpha _{3})
$. The superindices and subindices ($\sigma $) indicate whether a given
expression corresponds to the spacelike ($+$) or to the timelike ($-$) case.

Since the right hand side of (\ref{TPF}) is actually a correlator of a free
theory, we can easily write it down using the Wick contractions of the
exponential fields. The free field propagator in this case is $\left\langle
\varphi (z_{i})\varphi (z_{j})\right\rangle =-\sigma \log |z_{i}-z_{j}|$, so
that%
\begin{equation*}
C_{b}^{(\sigma )}(\alpha _{1},\alpha _{2},\alpha _{3})=\Gamma (-s_{\sigma
})\mu ^{s_{\sigma }}b^{-1}\tprod\nolimits_{l=1}^{s_{\sigma }}\int_{\mathbb{C}%
}d^{2}w_{l}\tprod\nolimits_{r=1}^{s_{\sigma }}|w_{r}|^{-4\sigma b\alpha
_{1}}|1-w_{r}|^{-4\sigma b\alpha _{2}}\tprod\nolimits_{t^{\prime
}<t}^{s_{\sigma }}|w_{t^{\prime }}-w_{t}|^{-4\sigma b^{2}}.
\end{equation*}%
This multiple integral of the Selberg type can be solved explicitly for
generic $s_{\pm }\in \mathbb{Z}_{>0}$. This was done by Dotsenko and Fateev
in the context of the Minimal Models in Ref. \cite{DF}. The result adapted
to Liouville theory reads%
\begin{equation}
C_{b}^{(\sigma )}(\alpha _{1},\alpha _{2},\alpha _{3})=(-1)^{s_{\sigma
}}\Gamma (-s_{\sigma })\Gamma (1+s_{\sigma })b^{-1}(\pi \mu b^{4}\gamma
(\sigma b^{2}))^{s_{\sigma }}\frac{\tprod\nolimits_{r=1}^{s_{\sigma }}\gamma
(-\sigma rb^{2})}{\tprod\nolimits_{i=1}^{3}\tprod\nolimits_{t=1}^{s_{\sigma
}}\gamma (2\sigma b\alpha _{i}+\sigma (t-1)b^{2})}  \label{arriba}
\end{equation}%
where $\gamma (x)=\Gamma (x)/\Gamma (1-x)$. The overall factor $\Gamma
(-s_{\pm })$ in the expressions above comes from the integration of the
zero-mode over the non-compact target space, and it is interpreted as in 
\cite{dFK} (see the discussion around Eq. (2.10) therein.) Notice that the
factor $\Gamma (-s_{\pm })$ would in principle introduce a divergence
provided $s_{\pm }\in \mathbb{Z}_{\geq 0}$. However, as we will see below,
in the spacelike case $\sigma =+1$ such divergent factor nicely cancels out
with a contribution coming from the products in (\ref{arriba}). The timelike
model, in contrast, exhibits a special feature and the divergence has to be
extracted.

Using the property $\gamma (x)=\gamma ^{-1}(1-x)$, we can write the products
appearing on the right hand side of (\ref{arriba}) as follows%
\begin{equation}
\tprod\nolimits_{l=1}^{s_{\sigma }}\gamma (-\sigma
lb^{2})\tprod\nolimits_{i=1}^{3}\left( \frac{\tprod\nolimits_{r=1}^{\beta
_{i}^{\sigma }}\gamma (rb^{2})}{\tprod\nolimits_{t=1}^{\beta _{i}^{\sigma
}+s_{\sigma }}\gamma (tb^{2})}\right) ^{\sigma },  \label{torbellino}
\end{equation}%
with $\beta _{i}^{\pm }=2\alpha _{i}b^{-1}-1+(1\mp 1)b^{-2}/2$. Here we
assumed $2\alpha _{i}b^{-1}\in \mathbb{Z}_{>0}$ and $b^{-2}\in \mathbb{Z}%
_{>0}$ to work out the products; however, we will eventually extend the
expression to generic values of $\alpha _{i}$ and $b$. To do this, we
consider the following expression%
\begin{equation}
\tprod\nolimits_{r=1}^{n}\gamma (rb^{2})=\frac{\Upsilon _{b}(nb+b)}{\Upsilon
_{b}(b)}b^{n(b^{2}(n+1)-1)},  \label{PUpsilon}
\end{equation}%
where $\Upsilon _{b}(x)$ is the special function introduced in \cite{ZZ},
which admits to be written in terms of the Barnes' double $\Gamma $%
-functions $\Gamma _{2}(x|y)$ \cite{Barnes} as follows%
\begin{equation*}
\Upsilon _{b}(x)\equiv \Gamma _{2}^{-1}(x|b,b^{-1})\Gamma
_{2}^{-1}(b+b^{-1}-x|b,b^{-1}),
\end{equation*}%
with the definition%
\begin{equation*}
\log \Gamma _{2}(x|y_{1},y_{2})=\lim_{\varepsilon \rightarrow 0}\frac{%
\partial }{\partial \varepsilon }\tsum\nolimits_{k_{1},k_{2}\in \mathbb{Z}%
_{\geq 0}}(x+k_{1}y_{1}+k_{2}y_{2})^{-\varepsilon };
\end{equation*}%
see also (\ref{integral}) below.

Notice that while Eq. (\ref{PUpsilon}) only makes sense for $n\in \mathbb{Z}%
_{>0}$, the expression on the right hand side is defined on a continuous
range. This is a crucial step in extending the Coulomb gas integral
expression to more general values of $\alpha _{i}$ and $b$. This extension
is clearly not unique; for instance, one could add a phase $e^{2\pi in}$ to
the right hand side of (\ref{PUpsilon}). In turn, the analytic continuation
of the integral formulas has to be regarded modulo certain type of
contributions.

Function $\Upsilon _{b}(x)$ presents simple zeros at $x=mb+nb^{-1}$ if $%
m,n\in \mathbb{Z}_{\leq 0}$ or $m,n\in \mathbb{Z}_{>0}$. Using the
properties of $\Gamma _{2}(x|y)$ it is possible to show that the $\Upsilon
_{b}(x)$ function satisfies the shift relations 
\begin{equation}
\Upsilon _{b}(x+b)=\gamma (bx)b^{1-2bx}\Upsilon _{b}(x),\qquad \Upsilon
_{b}(x+b^{-1})=\gamma (x/b)b^{-1+2x/b}\Upsilon _{b}(x)  \label{shifts}
\end{equation}%
as well as the inversion relations%
\begin{equation}
\Upsilon _{b}(x)=\Upsilon _{b^{-1}}(x),\qquad \Upsilon _{b}(x)=\Upsilon
_{b}(b+b^{-1}-x).  \label{inversions}
\end{equation}%
Expression (\ref{PUpsilon}) follows from iterating (\ref{shifts}).

Now, let us use the results above to compute the three-point function. Let
us first analyze the spacelike three-point function; namely, consider first
the case $\sigma =+1$. In this case, we can use (\ref{PUpsilon}) to write (%
\ref{torbellino}) as follows%
\begin{equation}
\tprod\nolimits_{l=1}^{s_{+}}\gamma (-lb^{2})\tprod\nolimits_{i=1}^{3}\frac{%
\tprod\nolimits_{r=1}^{2b^{-1}\alpha _{i}-1}\gamma (rb^{2})}{%
\tprod\nolimits_{t=1}^{2b^{-1}\alpha _{i}+s_{+}-1}\gamma (tb^{2})}=\frac{%
\Upsilon _{b}(Q_{+})b^{-2s(b^{2}+1)}}{\Upsilon
_{b}(\tsum\nolimits_{k=1}^{3}\alpha _{k}-Q_{+})}\tprod\nolimits_{i=1}^{3}%
\frac{\Upsilon _{b}(2\alpha _{i})}{\Upsilon
_{b}(\tsum\nolimits_{j=1}^{3}\alpha _{j}-2\alpha _{i})},  \label{tornadito}
\end{equation}%
which allows us to analytically extend the expression to values $2\alpha
_{i}/b\notin \mathbb{Z}$.

Then, using (\ref{PUpsilon}) and the functional properties (\ref{inversions}%
) we find the final expression%
\begin{equation}
C_{b}^{(+)}(\alpha _{1},\alpha _{2},\alpha _{3})=\left( \pi \mu \gamma
(b^{2})b^{2-2b^{2}}\right) ^{s_{+}}\frac{\Upsilon _{b}^{\prime }(0)}{%
\Upsilon _{b}(\tsum\nolimits_{k=1}^{3}\alpha _{k}-Q_{+})}\tprod%
\nolimits_{i=1}^{3}\frac{\Upsilon _{b}(2\alpha _{i})}{\Upsilon
_{b}(\tsum\nolimits_{j=1}^{3}\alpha _{j}-2\alpha _{i})}  \label{SC}
\end{equation}%
with $Q_{+}=b+b^{-1}$, $s_{+}=1+b^{-2}-b^{-1}\tsum\nolimits_{i=1}^{3}\alpha
_{i}$, and $\Upsilon _{b}^{\prime }(x)=\frac{\partial }{\partial x}\Upsilon
_{b}(x)$. To derive (\ref{SC}) we used the following integral expression for
the $\Upsilon _{b}(x)$ function,%
\begin{equation}
\log \Upsilon _{b}(x)=\int_{\mathbb{R}_{>0}}\frac{d\tau }{\tau }\left(
\left( \frac{b}{2}+\frac{1}{2b}-x\right) ^{2}e^{-\tau }-\frac{\sinh ^{2}((%
\frac{b}{2}+\frac{1}{2b}-x)\frac{\tau }{2})}{\sinh (\frac{b\tau }{2})\sinh (%
\frac{\tau }{2b})}\right)  \label{integral}
\end{equation}%
for $0<\func{Re}(x)<(b+b^{-1})/2$. In the boundary of this range we find
that the $\Upsilon _{b}(x)$ function behaves like $\Upsilon
_{b}(Q_{+})=\Upsilon _{b}(0)\sim \Upsilon _{b}^{\prime }(0)/\Gamma (0)$. The 
$\Gamma (0)$ appearing here cancels the one coming from $(-1)^{s_{+}}\Gamma
(-s_{+})$.

Expression (\ref{SC}) is the spacelike three-point function \cite{DO, ZZ}.
That is, the Coulomb gas approach based on the free field calculation
exactly reproduces the DOZZ\ formula, provided one analytically extends the
products standing in the integral formula (\ref{arriba}) using (\ref%
{PUpsilon}).

Now, let us proceed in the same way for the timelike case $\sigma =-1$. In
this case, we can write (\ref{torbellino}) as follows%
\begin{equation}
\tprod\nolimits_{r=1}^{s_{-}}\gamma (rb^{2})\tprod\nolimits_{i=1}^{3}\frac{%
\tprod\nolimits_{t=1}^{2\alpha _{i}b^{-1}+b^{-2}+s_{-}-1}\gamma (tb^{2})}{%
\tprod\nolimits_{r=1}^{2\alpha _{i}b^{-1}+b^{-2}-1}\gamma (rb^{2})}=\frac{%
\Upsilon _{b}(-\tsum\nolimits_{k=1}^{3}\alpha _{k}+Q_{-}+b)}{\Upsilon
_{b}(b)b^{2s(1-b^{2})}}\tprod\nolimits_{i=1}^{3}\frac{\Upsilon _{b}(2\alpha
_{i}-\tsum\nolimits_{j=1}^{3}\alpha _{j}+b)}{\Upsilon _{b}(b-2\alpha _{i})}.
\label{esta}
\end{equation}

From this we notice that, unlike the spacelike case, in which the
contribution $\Upsilon _{b}(0)\sim \Upsilon _{b}^{\prime }(0)/\Gamma (0)$
cancels the divergence coming from the factor $\Gamma (-s_{+})\sim
(-1)^{s_{+}}\Gamma (0)/\Gamma (1+s_{+})$, in the timelike case there is no $%
\Upsilon _{b}(0)$ factor coming from (\ref{esta}); instead, a finite
contribution $\Upsilon _{b}^{-1}(b)$ stands and the overall $\Gamma (-s_{-})$
factor then is not cancelled; in addition, a factor $(-1)^{s_{-}}$ survives.
The Coulomb gas calculation of the timelike case does yield a finite result
if, instead, we calculate the residues of the resonant correlators. This
amounts to extract the simple poles at $s_{-}\in \mathbb{Z}_{\geq 0}$ in the
momenta $\mathbb{C}$ plane. This is achieved by excluding the divergent
overall factor $\Gamma (0)$; recall $\Gamma (-s_{-})\sim (-1)^{s_{-}}\Gamma
(0)/s_{-}!$. The factor $1/s_{-}!=\Gamma ^{-1}(s_{-}+1)$ can alternatively
be thought of as a multiplicity factor coming from the permutation of the
screening operators $\mu e^{2b\varphi (w)}$ in the Feigin-Fuchs type
realization \cite{FF, DF} (see for instance Eq. (3.15) in \cite{ZZ}; see
also \cite{GL}). Then, using the properties of the $\Upsilon _{b}(x)$
functions, we find%
\begin{equation}
C_{b}^{(-)}(\alpha _{1},\alpha _{2},\alpha _{3})=\left( -\pi \mu \gamma
(-b^{2})b^{2+2b^{2}}\right) ^{s_{-}}\frac{\Upsilon
_{b}(-\tsum\nolimits_{k=1}^{3}\alpha _{k}+Q_{-}+b)}{b\Upsilon _{b}(b)}%
\tprod\nolimits_{i=1}^{3}\frac{\Upsilon _{b}(2\alpha
_{i}-\tsum\nolimits_{j=1}^{3}\alpha _{j}+b)}{\Upsilon _{b}(b-2\alpha _{i})}
\label{TC}
\end{equation}%
with $Q_{-}=b-b^{-1}$, $s_{-}=1-b^{-2}-b^{-1}\tsum\nolimits_{i=1}^{3}\alpha
_{i}$. The equality in (\ref{TC}) has to be understood after having computed
the residue of the expression.

Up to a phase $e^{-i\pi s_{-}}$, (\ref{TC}) turns out to reproduce exactly
the timelike three-point function recently discussed in \cite{HMW} and
originally proposed in \cite{Z} (notice that to translate our notation into
that used in \cite{HMW} one has to do $\alpha _{i}\rightarrow -\widehat{%
\alpha }_{i}$, $b\rightarrow \widehat{b}$, and $Q_{-}\rightarrow -\widehat{Q}
$.) About the phase, we already mentioned that a prescription to
analytically extend the products as in (\ref{PUpsilon}) is not sensitive to
phases like $e^{2\pi in}$. This is even more drastic in the case of multiple
products like those in (\ref{esta}), so that expression (\ref{TC}) has to be
regarded up to such a phase ambiguity. In this aspect, our result agrees
with that of \cite{KP}, which does not exhibit the phase $e^{-i\pi s_{-}}$
either. It is worthwhile emphasizing that, up to the phase, the expression
we obtained for $C_{b}^{(-)}(\alpha _{1},\alpha _{2},\alpha _{3})$
reproduces the normalization of Ref. \cite{HMW}, which differs from that in 
\cite{Z}; cf. Ref. \cite{05}. More precisely, in (\ref{TC}) we find exactly
the same factor $(-\mu \pi \gamma (-b^{2})b^{2+2b^{2}})^{s_{-}}$, as in \cite%
{HMW}, and find no additional $b$-dependent factors, in contrast with \cite%
{Z}. The only aspect about the normalization we may find puzzling is the
divergent $\Gamma (0)$ factor in which the spacelike and the timelike
differ. Despite one can easily keep track of such divergent factor through
the calculation, we do not find a simple way of explaining why it appears in
the timelike computation while it cancels out in the spacelike computation.
To try to understand this, we can check whether the same happens in the
partition function: The number of integrated screening operators in that
case would be $s_{\sigma }-3=-2+b^{2}$, instead of $s_{\sigma }$. This is
because to compute the genus-zero zero-point function we have to consider
the correlator with three local operators $e^{2b\varphi (z)}$ inserted at
fixed points, say $z_{1}=0$, $z_{2}=1$, and $z_{3}=\infty $. This stabilizes
the sphere compensating the volume of the conformal Killing group. Then,
assuming $s_{\sigma }\in \mathbb{Z}_{>3}$, the Liouville partition function
on the sphere topology reads%
\begin{equation*}
Z_{b}^{(\sigma )}=\frac{\mu ^{s_{\sigma }}}{b}\Gamma (-s_{\sigma })\Gamma
(s_{\sigma }-2)(-\pi \gamma (\sigma b^{2})b^{4})^{s_{\sigma
}-3}\prod_{r=1}^{s_{\sigma }-3}\frac{\gamma (-\sigma rb^{2})\gamma (\sigma
(s_{\sigma }+r-1)b^{2}-1)}{\gamma ^{2}(\sigma (1+r)b^{2})}.
\end{equation*}%
Noticing that $1-\sigma rb^{2}=\sigma b^{2}(s_{\sigma }-1-r)$ we can
rearrange the products of $\Gamma $-functions and eventually find for the
timelike case%
\begin{equation}
Z_{b}^{(-)}=\frac{(1+b^{2})\left( \pi \mu \gamma (-b^{2})\right) ^{Q_{-}/b}}{%
\gamma (-b^{2})\gamma (-b^{-2})\pi ^{3}Q_{-}},  \label{0pf}
\end{equation}%
recall $Q_{-}=b-b^{-1}$. A remarkable feature is that this expression is not
invariant under Liouville self-duality $b\rightarrow b^{-1}$ \cite{perturbed}%
. It should not be a surprise that (\ref{0pf}) agrees with the expression
obtained by replacing in the standard Liouville partition function as $%
Q_{+}\rightarrow Q_{-}$ and $b^{2}\rightarrow -b^{2}$. This is because,
unlike what happens with the three-point function, both the zero- and the
two-point function admit a natural analytic continuation to negative values
of $b^{2}$.

It is interesting to compare the three-point functions (\ref{SC}) and (\ref%
{TC}). It was early noticed in \cite{Z} and \cite{05} for the case of
Minimal Gravity that the timelike Liouville structure constants are not the
naive analytic continuation of their spacelike analogues. In fact, as
emphasized in \cite{Z}, contrary to one's expectation, the timelike
structure constants turn out to be, roughly speaking, the \textit{inverse}
of spacelike structure constants, in the sense that the product of both
timelike and spacelike quantities yields a remarkably simple factorized
expression. This peculiar relation between timelike and spacelike structure
constants is particularly expressed by the fact that the dependences on the $%
\Upsilon _{b}(x)$ functions in (\ref{SC}) and in (\ref{TC}) are, \textit{%
mutatis mutandis}, inverse of the other. This intriguing feature is nicely
explained in the Coulomb gas calculation as it directly follows from the
property $\gamma (x)=\gamma ^{-1}(1-x)$ of the $\Gamma $-functions. Exactly
the same happens in string theory on AdS$_{3}\times \mathbb{S}^{3}\times 
\mathbb{T}^{4}$, where the three-point functions of chiral states take a
remarkably simple expression due to surprising cancellations of $\Upsilon
_{b}(x)$ functions that take place between the $\mathbb{H}_{+}^{3}$ and the $%
\mathbb{S}^{3}$ pieces \cite{DP, GK}; in \cite{GN} such result was also
reproduced in the Coulomb gas approach. Here, this approach led us to\
reproduce the formula of \cite{HMW} for the timelike Liouville three-point
function. The fact that the analytic extension of the integral formulas
standing in the Coulomb gas calculation yields the correct answer both for
the spacelike and for the timelike Liouville structure constants is notable
because, as already mentioned, the latter are not simply obtained by
analytic continuation from the former.%
\begin{equation*}
\end{equation*}%
This work has been supported by NSF-CONICET cooperation grant, and by ANPCyT
and CONICET through grants PICT\ and PIP, respectively. The author thanks
Daniel Harlow, V\"{o}lker Schomerus, and Edward Witten for correspondence,
and thanks Matt Kleban and Massimo Porrati for conversations. The author
specially thanks Valya Petkova for addressing his attention to references 
\cite{05, KP, KP2}; in particular, in \cite{KP} the Coulomb gas calculation
for the CFT with $c<1$ was already discussed. He also thanks the Center for
Cosmology and Particle Physics of New York University for the hospitality
during his stay, where this work was done.

\end{document}